\documentclass{article}

\usepackage{amsmath}
\usepackage{amssymb}
\usepackage{amsfonts}
\usepackage{bm}
\usepackage{multicol}
\usepackage[dvips,xdvi]{graphicx}

\newlength{\refwidth}
\begin{document}


\title{\bf Information-Theoretical Approach to Relaxation Time Distribution in Rheology: 
Log-Normal Relaxation Spectrum Model}
\author{Takashi Uneyama\\
\\
Department of Materials Physics, Graduate School
 of Engineering, \\
Nagoya University, \\
Furo-cho, Chikusa, Nagoya 464-8603, Japan}

\maketitle

\begin{abstract}
The relaxation modulus of a viscoelastic fluid can be decomposed into multiple Maxwell models and
characterized by the relaxation spectrum for the relaxation time.
It is empirically known that the logarithmic relaxation time is
useful to express the relaxation spectrum. We use information geometry
to analyze the relaxation modulus and shown that the logarithmic
relaxation time is the most natural variable for the relaxation spectrum.
Then we use information theory to estimate the most probable functional
form for the relaxation spectrum. We show that the log-normal distribution
is the information-theoretically most probable relaxation spectrum.
We analyze the properties of the log-normal relaxation spectrum model
and compare it with the fractional Maxwell model. The fractional Maxwell model
with a small power-law exponent can be approximated as the log-normal relaxation
spectrum model with a large standard deviation.
We also compare the log-normal relaxation spectrum model with experimental
linear viscoelasticity data
for a high-density polyethylene, both at melt and solid states.
\end{abstract}

\maketitle

%

\section{INTRODUCTION}
\label{introduction}

The linear rheological behavior of a material can be characterized
by the shear relaxation modulus $G(t)$, and the shear relaxation modulus
can be expressed as\cite{Ferry-book}:
\begin{equation}
 \label{relaxation_modulus_with_linear_relaxation_spectrum}
 G(t) = \int_{0}^{\infty} d\tau \, \exp(- t / \tau) h(\tau),
\end{equation}
where $h(\tau)$ is the relaxation spectrum. Here we have assumed that
the the material can flow at the long-time limit and thus $G(t)$ decays exponentially at $t \to \infty$.
Empirically, it is convenient to use the
relaxation spectrum for the logarithmic relaxation time\cite{Ferry-book,Song-HoltenAndersen-McKinley-2023}:
\begin{equation}
 \label{relaxation_modulus_with_logarithmic_relaxation_spectrum}
 G(t)  = \int_{-\infty}^{\infty} d(\ln \tau) \, \exp(- t / \tau) H(\ln \tau),
\end{equation}
with $H(\ln \tau) = \tau h(\tau)$.
Intuitively, we expect that $H(\ln \tau)$ is preferred because it is
expressed in the logarithmic scale: rheological properties are typically
well described in the logarithmic scale than in the linear scale.
Also, the relaxation time can be distributed in a very wide range,
and the linear scale is not convenient.
However, as far as the author knows, there is
no {\em theoretical} support for the use of $H(\ln \tau)$.
The first question considered in this work is whether we can theoretically
explain that the use of $H(\ln \tau)$ is most suitable or not.

When we use eq~\eqref{relaxation_modulus_with_linear_relaxation_spectrum} or
\eqref{relaxation_modulus_with_logarithmic_relaxation_spectrum} to
analyze rheological data, sometimes we need to employ some models for $h(\tau)$
or $H(\ln \tau)$. For example, the Baumgaertel-Schausberger-Winter
(BSW) spectrum model\cite{Baumgaertel-Schausberger-Winter-1990} may be used to analyze entangled polymer melts.
If we have some microscopic or mesoscopic molecular models for
target systems, we may be able to estimate the relaxation spectra
starting from molecular models.
However, if the information about target systems is very limited,
we cannot employ molecular models for specific systems. Instead,
we will require a general relaxation spectrum model which is
applicable to various systems. The second question considered in this
work is what is this general relaxation spectrum model.

To answer two questions raised above, in this work we conduct
information-theoretical analyses for the relaxation modulus.
We show that the relaxation spectrum for the logarithmic relaxation
time $H(\ln \tau)$ can be interpreted as the most natural form, from the viewpoint of information geometry\cite{Amari-book,Nielsen-2020}.
(The meaning of ``natural'' used here will be explained later.)
We also show that the information-theoretically natural relaxation
spectrum is the Gaussian distribution for the logarithmic relaxation
time. It corresponds to the log-normal relaxation spectrum model\cite{Song-HoltenAndersen-McKinley-2023}.
We analyze the properties of the log-normal relaxation spectrum model
in detail. In some aspects, the log-normal relaxation spectrum model
has similar properties to the fractional Maxwell model\cite{Mainardi-book,Song-HoltenAndersen-McKinley-2023}. We compare two
models and show the similarities and differences between them.
Then we apply the log-normal
relaxation spectrum model to experimental shear relaxation modulus data
of a high-density polyethylene.
The terminal relaxation behavior of an entangled polydisperse
polymer melt can be reasonably reproduced by the log-normal relaxation spectrum
model. The storage and loss moduli of a crystalline polymer solid
can be also reproduced by the log-normal relaxation spectrum model.

\section{THEORY}
\label{theory}

\subsection{Information Geometry for Relaxation Spectrum}
\label{information_geometry_for_relaxation_spectrum} 

The relaxation modulus of the (single) Maxwell model with the relaxation time $\tau$ and
the relaxation intensity (or the limiting modulus) $g$ is expressed as
\begin{equation}
 \label{relaxation_modulus_single_maxwell_model}
 G_{M}(t|\tau) = g \exp(-t / \tau).
\end{equation}
If we define the normalized relaxation modulus as $\Phi_{M}(t|\tau) = G_{M}(t|\tau) / g = \exp(- t / \tau)$,
we have $\Phi_{M}(0|\tau) = 1$ and $\partial \Phi_{M}(t|\tau) / \partial t < 0$ 
($\Phi_{M}(t|\tau)$ is a monotonically decreasing function of $t$).
Therefore, we may interpret $\Phi_{M}(t|\tau)$ as the survival function\cite{vanKampen-book} of the stress.
This interpretation can be also supported by a microscopic molecular model.
According to the linear response theory\cite{Evans-Morriss-book},
the relaxation modulus can be related to the auto-correlation function of the stress fluctuation.
If we normalize the relaxation modulus, it becomes the normalized correlation function
which may be interpreted as the survival function.
By taking the derivative of the survival function, we have the
probability density:
\begin{equation}
 \label{probability_density_single_maxwell_model}
 P_{M}(t | \tau) = - \frac{\partial \Phi_{M}(t|\tau)}{\partial t} 
  = \frac{1}{\tau} \exp(-t / \tau).
\end{equation}
Eq~\eqref{probability_density_single_maxwell_model} can be interpreted
as the probability density for the ``life time'' of the stress.

In general, the relaxation modulus is expressed as the superposition
of Maxwell models with various relaxation times (the generalized
Maxwell model), as eq~\eqref{relaxation_modulus_with_linear_relaxation_spectrum}.
Then the probability density for the life time becomes
\begin{equation}
 \label{probability_density_with_linear_relaxation_spectrum}
 P(t) = - \frac{1}{G_{0}} \frac{\partial G(t)}{\partial t} = \int_{0}^{\infty} d\tau \, P_{M}(t | \tau) \tilde{h}(\tau),
\end{equation}
with the total relaxation intensity $G_{0}$ and the normalized relaxation spectrum $\tilde{h}(\tau)$ defined as
\begin{align}
 \label{total_relaxation_intensity}
 G_{0} & = G(0) = \int_{0}^{\infty} d\tau \, h(\tau), \\
 \label{normalized_linear_relaxation_spectrum}
 \tilde{h}(\tau) & = \frac{1}{G_{0}} h(\tau).
\end{align}
$\tilde{h}(\tau)$ is normalized and thus it can be interpreted as
the probability density for the relaxation time $\tau$.
Eq~\eqref{probability_density_with_linear_relaxation_spectrum} may be interpreted as the combination
of two probability densities, $P_{M}(t|\tau)$ and $\tilde{h}(\tau)$, and
such a combination is sometimes
called superstatistics\cite{Beck-Cohen-2003}.

If we introduce a (nonlinear) variable transform from $\tau$ to $\xi$,
we will have the probability density $P(t)$ expressed in terms of the
probability density for $\xi$. 
Any variable transform gives the same probability density $P(t)$.
But the superstatistical meaning depends on $\xi$. In general,
the parameter space for $\xi$ is not a Euclidean space but a Riemannian space,
and the metric which characterizes it is not constant.
We consider that the most ``natural'' variable transform is that gives
constant metric.

Here we utilize the theory of information geometry\cite{Amari-book,Nielsen-2020}
to seek the natural variable transform.
In information geometry,
we can quantitatively measure the ``distance'' between different probability
densities and analyze the geometrical properties for probability densities.
They will be useful to select the natural variable transform
among various candidates.

We consider two Maxwell models with different relaxation times.
The distance between two probability densities with different relaxation times
can be calculated based on the probability density \eqref{probability_density_single_maxwell_model}.
We assume that two relaxation times for two probability densities are sufficiently close: 
$\tau$ and $\tau + d\tau$ ($d\tau$ is assumed to be sufficiently small).
According to information geometry\cite{Amari-book,Nielsen-2020}, the distance between these probability densities, $dL$, is given as
\begin{equation}
 \label{informational_distance_single_maxwell_model}
 dL^{2} = I(\tau) d\tau^{2},
\end{equation}
where $I(\tau)$ is the Fisher information defined as:
\begin{equation}
 \label{fisher_information_single_maxwell_model}
   I(\tau)  = \int_{0}^{\infty} dt \, P_{M}(t | \tau) 
  \left[ \frac{\partial \ln P_{M}(t | \tau)}{\partial \tau} \right]^{2} 
   = \frac{1}{\tau^{2}}.
\end{equation}
Intuitively, the Fisher information \eqref{fisher_information_single_maxwell_model} characterizes the degree
of change of $P_{\text{M}}(t|\tau)$ to the change of its parameter $\tau$.
(The Fisher information was originally introduced for an optimization problem\cite{Frieden-book}, 
but currently it is also utilized in several different fields such as information geometry.)
By combining eqs~\eqref{informational_distance_single_maxwell_model}
and \eqref{fisher_information_single_maxwell_model}, we have the
distance between two probability densities with relaxation times $\tau'$ and $\tau''$:
\begin{equation}
 \label{distance_between_two_single_maxwell_models}
 L(\tau'',\tau') = \int_{\tau'}^{\tau'' } dL = 
 \int_{\tau'}^{\tau''} \sqrt{ \frac{1}{\tau^{2}} d\tau^{2}  }
  = |\ln \tau' - \ln \tau''|.
\end{equation}

Eq~\eqref{distance_between_two_single_maxwell_models} means that the
distance between two probability densities with $\tau'$ and $\tau''$
is {\em not} $|\tau' - \tau''|$ (which corresponds to the simple Euclidean distance).
This is not surprising because the information geometry states that
the space for a family of probability densities generally becomes a Riemannian space.
The Fisher information $I(\tau)$ 
in eq~\eqref{informational_distance_single_maxwell_model} can be interpreted as the metric
in the Riemannian geometry\cite{Dirac-book}.
By using a variable transform, a one-dimensional Riemannian space 
can be transformed into a one-dimensional Euclidean space.
(This is because the Riemann curvature is trivially zero in a one-dimensional space\cite{Dirac-book}.)
We employ $\xi = \ln (\tau / \tau_{0})$ instead of $\tau$. Here, $\tau_{0}$ is a
characteristic relaxation time of the target system.
The distance between $\xi' = \ln (\tau' / \tau_{0})$ and $\xi'' = \ln (\tau'' / \tau_{0})$ simply becomes
$L(\xi'',\xi') = |\xi' - \xi''|$, which is the Euclidean distance.
The Fisher information for $\xi$ becomes unity, as expected:
\begin{equation}
 \label{fisher_information_single_maxwell_model_xi}
   I'(\xi) = \int_{0}^{\infty} dt \, P_{M}(t | \tau_{0} e^{\xi}) 
  \left[ \frac{\partial \ln P_{M}(t | \tau_{0} e^{\xi})}{\partial \xi} \right]^{2} = 1.
\end{equation}
With this new variable, we have
\begin{equation}
 P(t) = \int_{-\infty}^{\infty} d\xi \, P_{M}(t | \tau_{0} e^{\xi}) \tilde{H}(\xi),
\end{equation}
where
\begin{equation}
 \tilde{H}(\xi) = \tau_{0} e^{\xi} \tilde{h}(\tau_{0} e^{\xi})
\end{equation}
is the normalized relaxation spectrum. $\tilde{H}(\xi)$ can be interpreted as
the probability density for $\xi$.

From the viewpoint of information geometry, $\xi = \ln (\tau / \tau_{0})$ is the most ``natural'' parameter
which characterizes the probability density \eqref{probability_density_single_maxwell_model}.
Therefore, the information-theoretically most natural expression for
the relaxation modulus $G(t)$ is
\begin{equation}
 \label{relaxation_modulus_xi}
  G(t) = \int_{-\infty}^{\infty} d\xi \, \exp(- t / \tau_{0} e^{-\xi}) H(\xi),
\end{equation}
with 
\begin{equation}
 H(\xi) = G_{0} \tilde{H}(\xi) = \tau_{0} e^{\xi} h(\tau_{0} e^{\xi}). 
\end{equation}
Eq~\eqref{relaxation_modulus_xi} coincides with eq~\eqref{relaxation_modulus_with_logarithmic_relaxation_spectrum}
if we set $\tau_{0} = 1$. ($\tau_{0}$ is the characteristic
relaxation time and can be taken rather arbitrarily.)
We conclude that the relaxation spectrum for the logarithmic
relaxation time, $H(\xi)$, is {\em the most natural and suitable
expression}, from the view point of information geometry.
This is the answer to the first question.

\subsection{Most Probable Distribution for Relaxation Time}
\label{most_probable_distribution_for_relaxation_time}

If $\xi = \ln (\tau / \tau_{0})$ is used instead of $\tau$,
we can treat the space for the family of $P_{M}(t|\tau_{0} e^{\xi})$ 
as just the one dimensional Euclidean space.
We use this property to determine the most probable functional form
for the relaxation spectrum.
As we mentioned, $\tilde{H}(\xi)$ is normalized and thus can be interpreted
as the probability density for $\xi$.

In this subsection, we derive the information-theoretically most probable probability
density for $\xi$. According to information theory, the most probable
probability density is given as the probability density which maximizes
the Shannon entropy. The Shannon entropy is given as a functional of the
probability density $\tilde{H}(\xi)$\cite{Shannon-Weaver-book}:
\begin{equation}
 \label{shannon_entropy_xi}
  \mathcal{S}[\tilde{H}] = - \int_{-\infty}^{\infty} d\xi \, \tilde{H}(\xi)
  \ln \tilde{H}(\xi).
\end{equation}
Here, we should stress that the Shannon entropy is well-defined only
in the Euclidean space. Thus we should {\em not} utilize $\tau$ and $\tilde{h}(\tau)$
instead of $\xi$ and $\tilde{H}(\xi)$.

The probability density which maximizes eq~\eqref{shannon_entropy_xi}
under some constraints can be interpreted as the information-theoretically most probable distribution.
Since $\tilde{H}(\xi)$ can be interpreted as
the probability density, it should be normalized and we need to introduce the following constraint:
\begin{equation}
 \label{normalization_constraint}
  \int_{-\infty}^{\infty} d\xi \, \tilde{H}(\xi) = 1.
\end{equation}
In addition, some moments of $\xi$ need to be constrained.
The first moment of $\xi$ is the mean, but we do not need to constrain
it. We have already introduced the characteristic relaxation time $\tau_{0}$
and it can be used to control the mean value of $\ln \tau$.
The second moment of $\xi$ is the variance (or the squared standard
deviation). Thus we employ the following constraint:
\begin{equation}
 \label{standard_deviation_constraint}
  \int_{-\infty}^{\infty} d\xi \, \xi^{2} \tilde{H}(\xi) = s^{2},
\end{equation}
where $s \ge 0$ is the standard deviation of $\xi$.
We maximize the Shannon entropy \eqref{shannon_entropy_xi} with respect to
$\tilde{H}(\xi)$ under the constraints \eqref{normalization_constraint}
and \eqref{standard_deviation_constraint}. The maximization can be easily done
by using the Lagrange multiplier method. That is, we maximize
\begin{equation}
 \label{shannon_entropy_xi_with_constraints}
 \mathcal{S}'[\tilde{H}]
  = \mathcal{S}[\tilde{H}]
  + \lambda \int_{-\infty}^{\infty} d\xi \, \tilde{H}(\xi)
  + \mu \int_{-\infty}^{\infty} d\xi \, \xi^{2} \tilde{H}(\xi),
\end{equation}
instead of $\mathcal{S}[\tilde{H}]$. Here, $\lambda$ and $\mu$ are
Lagrange multipliers and they will be determined afterward, so that the constraints
\eqref{normalization_constraint} and \eqref{standard_deviation_constraint} are satisfied. The maximization of eq~\eqref{shannon_entropy_xi_with_constraints}
gives
\begin{equation}
 \label{shannon_entropy_xi_with_constraints_maximization}
  0  = \frac{\delta \mathcal{S}'[\tilde{H}]}{\delta \tilde{H}(\xi)} = 1 + \ln \tilde{H}(\xi) + \lambda + \mu \xi^{2}.
\end{equation}
In eq~\eqref{shannon_entropy_xi_with_constraints_maximization}, $\delta /\delta \tilde{H}(\xi)$ represents the
functional differential.
The solution of eq~\eqref{shannon_entropy_xi_with_constraints_maximization} is
the normal distribution:
\begin{equation}
 \label{relaxation_spectrum_xi_most_probable}
 \tilde{H}(\xi) = \frac{1}{\sqrt{2 \pi s^{2}}}
  \exp\left( - \frac{\xi^{2}}{2 s^{2}}\right),
\end{equation}
where we have determined the Lagrange multipliers as $\lambda = - 1 + (1 / 2) \ln (2 \pi s^{2})$
and $\mu = 1 / 2 \sigma^{2}$ from constraints \eqref{normalization_constraint} and \eqref{standard_deviation_constraint}, and substituted them into the expression of $\tilde{H}(\xi)$.

Eq~\eqref{relaxation_spectrum_xi_most_probable} is the information-theoretically
most probable relaxation spectrum under the constraints
\eqref{normalization_constraint} and \eqref{standard_deviation_constraint}.
If we introduce some additional constraints, we have different 
most probable relaxation spectra. But the number of parameters increase
as we increase the number of constraints.
A model with the smallest number of constraints will be convenient.
If we remove the constraint for the standard deviation, eq~\eqref{standard_deviation_constraint}, the resulting
distribution will be that with the smaller number of parameters.
However, this is not possible. Without the constraint for the standard deviation,
the maximization condition becomes $0 = 1 + \ln \tilde{H}(\xi) + \lambda$
instead of eq~\eqref{shannon_entropy_xi_with_constraints_maximization}.
This gives the constant probability density:
$\tilde{H}(\xi) = (\text{const}.)$. But $\xi$ is defined
in the range $-\infty < \xi < \infty$ and the constant probability density
cannot be realized.
Thus we conclude that the normal distribution \eqref{relaxation_spectrum_xi_most_probable} is the most probable functional form
with the smallest number of parameters.

The unnormalized relaxation spectra $H(\xi)$ and $h(\tau)$ which correspond to eq~\eqref{relaxation_spectrum_xi_most_probable}
become
\begin{align}
 \label{logarithmic_relaxation_spectrum_most_probable}
 H(\xi) & = \frac{G_{0}}{\sqrt{2 \pi s^{2}}}
 \exp\left( - \frac{\xi^{2}}{2 s^{2}}\right), \\
 \label{linear_relaxation_spectrum_most_probable}
 h(\tau) & = \frac{G_{0}}{\sqrt{2 \pi s^{2}} \tau}
  \exp\left[ - \frac{(\ln \tau - \ln \tau_{0})^{2}}{2 s^{2}}\right].
\end{align}
Eq~\eqref{linear_relaxation_spectrum_most_probable} is the log-normal
probability density. Therefore, we call the linear viscoelasticity model
with eq~\eqref{logarithmic_relaxation_spectrum_most_probable} or eq~\eqref{linear_relaxation_spectrum_most_probable} as the
log-normal relaxation spectrum model\cite{Song-HoltenAndersen-McKinley-2023}.
The answer to the second question is that we can use
the log-normal relaxation spectrum model if the information about
the target systems is very limited.
It is the {\em information-theoretically most probable} relaxation spectrum model.
The log-normal relaxation spectrum model
has three parameters: the total relaxation intensity $G_{0}$,
the characteristic relaxation time $\tau_{0}$,
and the standard deviation $s$.
$G_{0}$ and $\tau_{0}$ simply determine the stress and time scales,
and the shape of the relaxation modulus depends only on $s$.

\subsection{Properties of Log-Normal Relaxation Spectrum Model}
\label{properties_of_log-normal_relaxation_spectrum_model}

Although
the use of the log-normal relaxation spectrum model
has been reported in literature\cite{Fulchiron-Michel-Verney-Roustant-1995,Grindy-Learsch-Mozhdehi-Cheng-Barrett-Guan-Messersmith-HoltenAndersen-2015,Drozdov-Christiansen-2021},
it is not that popular.
Thus its properties are not well-known, compared with other empirical models.
Here we study some properties of the log-normal
relaxation time spectrum model.
The relaxation modulus is given by substituting 
eq~\eqref{logarithmic_relaxation_spectrum_most_probable} into
eq~\eqref{relaxation_modulus_xi}.
The storage and loss moduli are calculated
from the relaxation spectrum as
\begin{align}
 \label{storage_modulus_log_normal_relaxation_spectrum}
 G'(\omega) & = \int_{-\infty}^{\infty} d\xi \, \frac{(\tau_{0} e^{\xi} \omega)^{2}}{1 + (\tau_{0} e^{\xi} \omega)^{2}} H(\xi), \\
 \label{loss_modulus_log_normal_relaxation_spectrum}
 G''(\omega) & = \int_{-\infty}^{\infty} d\xi \, \frac{\tau_{0} e^{\xi} \omega}{1 + (\tau_{0} e^{\xi} \omega)^{2}} H(\xi),
\end{align}
with $H(\xi)$ by eq~\eqref{logarithmic_relaxation_spectrum_most_probable}.
At the limit of $s \to 0$, $H(\xi)$ approaches to the Dirac delta function:
$H(\xi) \to G_{0} \delta(\xi)$. Thus the log-normal relaxation model reduces to
the Maxwell model at $s = 0$.

\begin{figure}[tb]
\begin{center}
 \includegraphics[width=1.\refwidth]{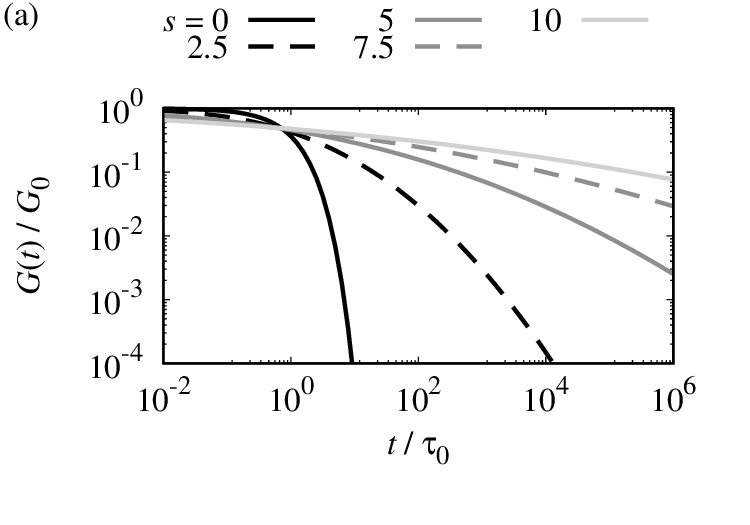} \\
 \includegraphics[width=1.\refwidth]{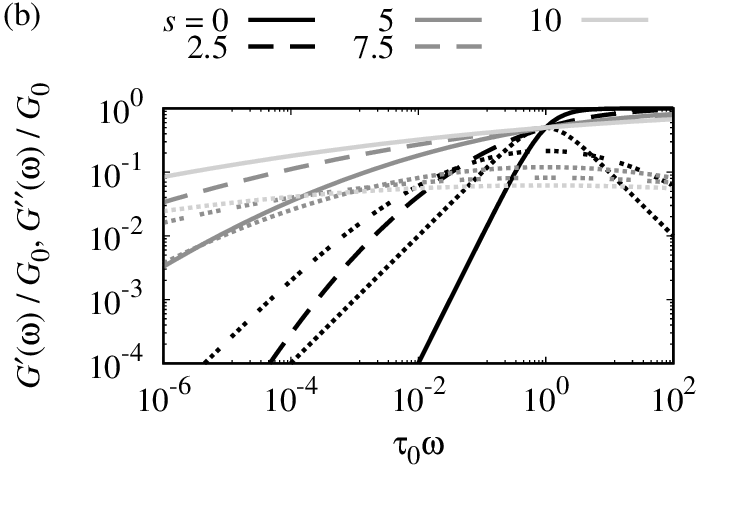}%
\end{center}
\caption{
 (a) Relaxation moduli $G(t)$ and (b) storage and loss moduli $G'(\omega)$ and $G''(\omega)$
 of the log-normal relaxation spectrum model with $s = 0, 2.5, 5, 7.5$, and $10$.
 In (b), solid and dashed curves indicate $G'(\omega)$ whereas
 dotted and dot-dashed curves indicate $G''(\omega)$.
 In the case of $s = 0$, the log-normal relaxation spectrum model reduces to the Maxwell model.
 \label{relaxaton_modulus_and_complex_modulus_log_normal}}
\end{figure}

We numerically calculate $G(t)$, $G'(\omega)$, and $G''(\omega)$ of the log-normal relaxation spectrum model.
The double exponential formula is used to perform the numerical integration method\cite{Takahashi-Mori-1974,Ooura-program}.
Fig.~\ref{relaxaton_modulus_and_complex_modulus_log_normal} shows
numerically calculated $G(t)$, $G'(\omega)$, and $G''(\omega)$ for
$s = 2.5, 5, 7.5,$ and $10$. For comparison, the data for $s = 0$ (which
corresponds to the Maxwell model) are also shown.
As $s$ increases, we observe that moduli become broad.
$G(t)$ with $s = 10$ do not decay exponentially at least in the plotted range ($t / \tau_{0} \le 10^{6}$).
Also, we cannot observe the terminal region for $G'(\omega)$ and $G''(\omega)$
with $s = 10$ in the plotted range ($\tau_{0} \omega \ge 10^{-6}$).
However, the log-normal relaxation model exhibits the terminal behavior
at the sufficiently low frequency region for any $s$, as we will show.
The loss modulus $G''(\omega)$ in Fig.~\ref{relaxaton_modulus_and_complex_modulus_log_normal}(b)
seems symmetric around its peak at $\tau_{0} \omega = 1$.
This symmetry can be derived straightforwardly.
From eq~\eqref{logarithmic_relaxation_spectrum_most_probable}, $H(\xi)$ is
symmetric: $H(\xi) = H(-\xi)$. If we introduce a variable transform $\xi' = - \xi$,
eq~\eqref{loss_modulus_log_normal_relaxation_spectrum} gives
\begin{equation}
 \label{loss_modulus_log_normal_relaxation_spectrum_symmetry}
\begin{split}
  G''(\omega) & = \int_{-\infty}^{\infty} d\xi' \, \frac{\tau_{0} e^{-\xi'} \omega}{1 + (\tau_{0} e^{-\xi'} \omega)^{2}} H(-\xi') \\
 & = \int_{-\infty}^{\infty} d\xi' \, \frac{\tau_{0} e^{\xi'} (\tau_{0}^{-2} \omega^{-1})}{1 + [\tau_{0} e^{\xi'} (\tau_{0}^{-2} \omega^{-1})]^{2}} H(\xi') \\
 & = G''(\tau_{0}^{- 2} \omega^{-1}).
\end{split}
\end{equation}

\begin{figure}[tb]
\begin{center}
 \includegraphics[width=1.\refwidth]{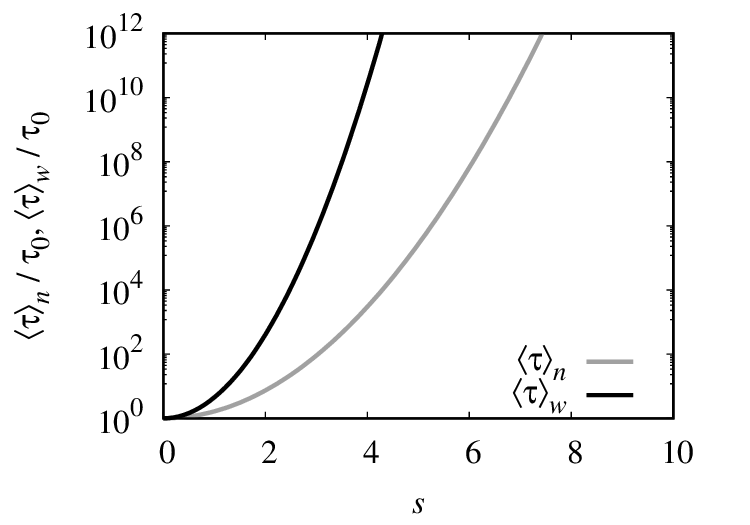}%
\end{center}
\caption{First- and second-moment average relaxation times 
 $\langle \tau \rangle_{n}$ and $\langle \tau \rangle_{w}$ for
 the log-normal relaxation spectrum model.
 \label{relaxaton_time_log_normal}}
\end{figure}

The average of $\tau^{p}$ (with $p$ being an integer) is calculated as
\begin{equation}
 \langle \tau^{p} \rangle = \int_{-\infty}^{\infty} d\xi \, (\tau_{0} e^{\xi})^{p} \tilde{H}(\xi)
  = \tau_{0}^{p} \exp(p^{2} s^{2} / 2).
\end{equation}
Then we have the following first- and second-moment average relaxation times:
\begin{align}
 \label{first_moment_average_relaxation_time}
 \langle \tau \rangle_{n}
 & = \langle \tau \rangle 
  = \tau_{0} \exp(s^{2} / 2), \\
 \label{second_moment_average_relaxation_time}
 \langle \tau \rangle_{w}
 & = {\langle \tau^{2} \rangle} / {\langle \tau \rangle}
  = \tau_{0} \exp(3 s^{2} / 2).
\end{align}
Thus both $\langle \tau \rangle_{n}$ and $\langle \tau \rangle_{w}$
are finite. This is in contrast to some empirical models such as
the fractional Maxwell model\cite{Mainardi-book,Song-HoltenAndersen-McKinley-2023}.
Fig.~\ref{relaxaton_time_log_normal} shows $\langle \tau \rangle_{n}$ and $\langle \tau \rangle_{w}$
by eqs~\eqref{first_moment_average_relaxation_time} and \eqref{second_moment_average_relaxation_time}.
We can observe that both relaxation times increase rapidly as $s$ increases.
If $s$ is relatively large, practically we will not be able to
observe the relaxation.

Eqs~\eqref{first_moment_average_relaxation_time} and
\eqref{second_moment_average_relaxation_time} mean that
the terminal behaviors are observed for the storage and loss moduli
at the low-frequency region.
For $\tau_{0} \omega \ll 1$, we have
\begin{align}
 \label{storage_modulus_log_normal_low_frequency}
 G'(\omega) & \approx G_{0} \langle \tau^{2} \rangle \omega^{2}
 = G_{0} \tau_{0}^{2} \exp(2 s^{2}) \, \omega^{2}, \\
 \label{loss_modulus_log_normal_low_frequency}
 G''(\omega) & \approx G_{0} \langle \tau \rangle \omega = G_{0} \tau_{0} \exp(s^{2} / 2) \, \omega.
\end{align}
From eq~\eqref{loss_modulus_log_normal_low_frequency}, 
the zero-shear viscosity is
\begin{equation}
 \eta_{0} = G_{0} \tau_{0} \exp(s^{2} / 2).
\end{equation}
For $\tau_{0} \omega \gg 1$, we have
\begin{align}
 \label{storage_modulus_log_normal_high_frequency}
 G'(\omega) & \approx G_{0}, \\
 \label{loss_modulus_log_normal_high_frequency}
 G''(\omega) & \approx \frac{G_{0} \langle \tau \rangle}{\tau_{0}^{2} \omega} = \frac{G_{0} \exp(s^{2} / 2)}{\tau_{0} \omega}.
\end{align}
Here, we have utilized eqs~\eqref{loss_modulus_log_normal_relaxation_spectrum_symmetry}
and \eqref{loss_modulus_log_normal_low_frequency} to derive eq~\eqref{loss_modulus_log_normal_high_frequency}.

\section{DISCUSSIONS}
\label{discussions}

\subsection{Comparison with Fractional Maxwell Model}
\label{comparison_with_fractional_maxwell_model}

One interesting property of the log-normal relaxation
spectrum model is that $\tilde{H}(\xi)$ is symmetric: $\tilde{H}(\xi) = \tilde{H}(-\xi)$.
In this section we compare the log-normal relaxation spectrum
model with the fractional Maxwell model\cite{Mainardi-book,Song-HoltenAndersen-McKinley-2023},
of which $\tilde{H}(\xi)$ has the same symmetry.
The fractional Maxwell model is often related to the fractional derivative model\cite{Mainardi-book,Urakawa-Nobukawa-Inoue-2023},
and it is equivalent to the Cole-Cole model for dielectric relaxation\cite{Cole-Cole-1941,Hilfer-2002,Urakawa-Nobukawa-Inoue-2023}.

The fractional Maxwell model can be defined as a simple form in the frequency domain.
The storage and loss moduli are given as
\begin{equation}
 \label{complex_modulus_fractional_maxwell}
  G'(\omega) + i G''(\omega) = G_{0} \frac{(i \tau_{0} \omega)^{\alpha}}{1 + (i \tau_{0} \omega)^{\alpha}},
\end{equation}
where $G_{0}$ is the total relaxation intensity, $\tau_{0}$ is the characteristic
relaxation time, and $0 < \alpha \le 1$ is the exponent.
(For $\alpha = 1$, eq~\eqref{complex_modulus_fractional_maxwell} reduces to the
Maxwell model~\eqref{relaxation_modulus_single_maxwell_model}.)
Eq~\eqref{complex_modulus_fractional_maxwell} can be rewritten as follows:
\begin{align}
 \label{storage_modulus_fractional_maxwell}
 G'(\omega) & = G_{0} \frac{(\tau_{0} \omega)^{2 \alpha} + \cos(\alpha \pi / 2) \, (\tau_{0} \omega)^{\alpha}}{[(\tau_{0} \omega)^{\alpha} + \cos(\alpha \pi / 2)]^{2} + \sin^{2}(\alpha \pi / 2)},\\
 \label{loss_modulus_fractional_maxwell}
 G''(\omega) & = G_{0} \frac{\sin(\alpha \pi / 2) \, (\tau_{0} \omega)^{\alpha} }{[(\tau_{0} \omega)^{\alpha} + \cos(\alpha \pi / 2)]^{2} + \sin^{2}(\alpha \pi / 2)}.
\end{align}
The fractional Maxwell model does not exhibit the terminal behavior.
For $\tau_{0} \omega \ll 1$, we have
\begin{align}
 \label{storage_modulus_fractional_maxwell_low_frequency}
 G'(\omega) & \approx G_{0} \cos(\alpha \pi / 2) \, (\tau_{0} \omega)^{\alpha}, \\
 \label{loss_modulus_fractional_maxwell_low_frequency}
 G''(\omega) & \approx G_{0} \sin(\alpha \pi / 2) \, (\tau_{0} \omega)^{\alpha} .
\end{align}
Thus the fractional Maxwell model behaves in the same way as the critical gel\cite{Winter-Chambon-1986}:
$G'(\omega) \propto G''(\omega) \propto \omega^{\alpha}$ and $\tan \delta(\omega) = \tan(\alpha \pi / 2)$.
This means that the fractional Maxwell model does not flow.
(The same critical-gel behavior is observed for the Havriliak-Negami model\cite{Kawasaki-Watanabe-Uneyama-2011}.)
The first- and second-moment average relaxation times and the zero-shear viscosity diverge.
This is in contrast to the log-normal relaxation spectrum model.

In the time domain, the relaxation modulus $G(t)$ of the fractional Maxwell model cannot be expressed
in terms of the elementary functions. $G(t)$ is expressed as
\begin{equation}
 \label{relaxation_modulus_fractional_maxwell}
 G(t) = G_{0} E_{\alpha}[- (t / \tau_{0})^{\alpha}],
\end{equation}
where $E_{\alpha}(x)$ is the Mittag-Leffler function\cite{Mainardi-book,Mainardi-2014}.
For $0 < \alpha < 1$ and $t \gg \tau_{0}$, eq~\eqref{relaxation_modulus_fractional_maxwell}
can be approximated as
\begin{equation}
 \label{relaxation_modulus_fractional_maxwell_long_time}
 G(t) \approx G_{0} \frac{ (t / \tau_{0})^{\alpha}}{\Gamma(1 - \alpha)},
\end{equation}
where $\Gamma(x)$ is the gamma function\cite{NIST-handbook}.
Thus the fractional Maxwell model exhibits the power-law tail: $G(t) \propto t^{-\alpha}$.
This is consistent with the lack of the terminal behavior in the storage
and loss moduli.

\begin{figure}[tb]
\begin{center}
 \includegraphics[width=1.\refwidth]{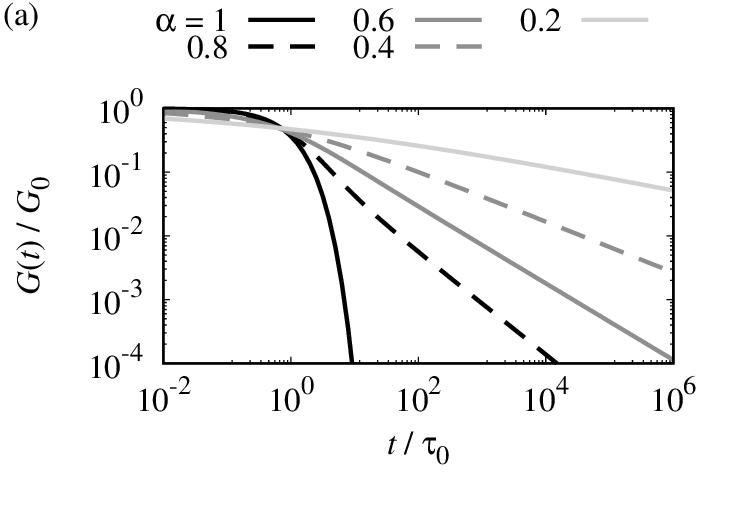} \\
 \includegraphics[width=1.\refwidth]{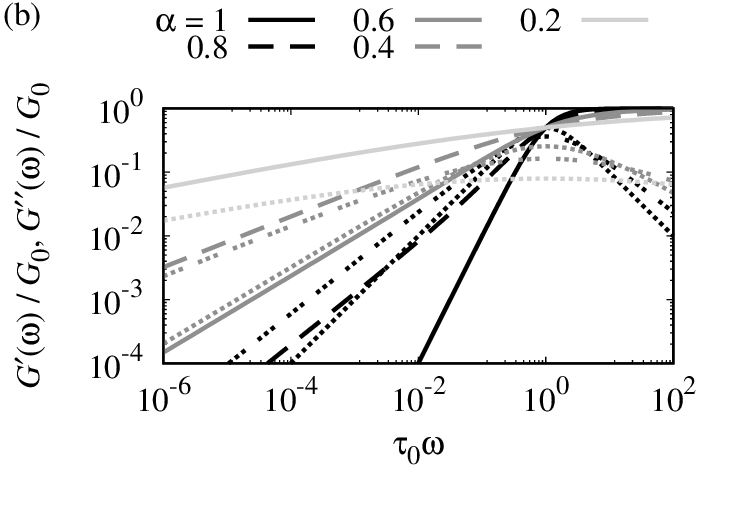}%
\end{center}
\caption{
 (a) Relaxation moduli $G(t)$ and (b) storage and loss moduli $G'(\omega)$ and $G''(\omega)$
 of the fractional Maxwell model with $\alpha = 0.2, 0.4, 0.6, 0.8,$ and $1$.
 $\alpha = 1$ corresponds to the Maxwell model.
 \label{relaxaton_modulus_and_complex_modulus_fractional_maxwell}}
\end{figure}

Fig.~\ref{relaxaton_modulus_and_complex_modulus_fractional_maxwell} shows
$G(t)$, $G'(\omega)$, and $G''(\omega)$ of the fractional Maxwell model
with $\alpha = 0.2, 0.4, 0.6, 0.8$, and $1$.
Here, the Mittag-Leffler function is numerically calculated\cite{Viet-program}.
We observe the power-law tail (eq~\eqref{relaxation_modulus_fractional_maxwell_long_time})
at the long-time region in Fig.~\ref{relaxaton_modulus_and_complex_modulus_fractional_maxwell}(a),
and the critical-gel behavior (eqs~\eqref{storage_modulus_fractional_maxwell_low_frequency} and \eqref{loss_modulus_fractional_maxwell_low_frequency})
at the low-frequency region in Fig.~\ref{relaxaton_modulus_and_complex_modulus_fractional_maxwell}(b),
except the case of $\alpha = 1$.
The loss modulus data in Fig.~\ref{relaxaton_modulus_and_complex_modulus_fractional_maxwell}
seem to have the symmetry $G''(\omega) = G''(\tau_{0}^{-2} \omega^{-1})$, which is
the same property as the log-normal relaxation spectrum model.
If we compare the data in Figs.~\ref{relaxaton_modulus_and_complex_modulus_log_normal}
and \ref{relaxaton_modulus_and_complex_modulus_fractional_maxwell},
we observe that the fractional Maxwell model with relatively small $\alpha$
(such as $\alpha = 0.2$)
exhibits similar trends to the log-normal relaxation spectrum model
with large $s$.

\begin{figure}[tb]
\begin{center}
 \includegraphics[width=1.\refwidth]{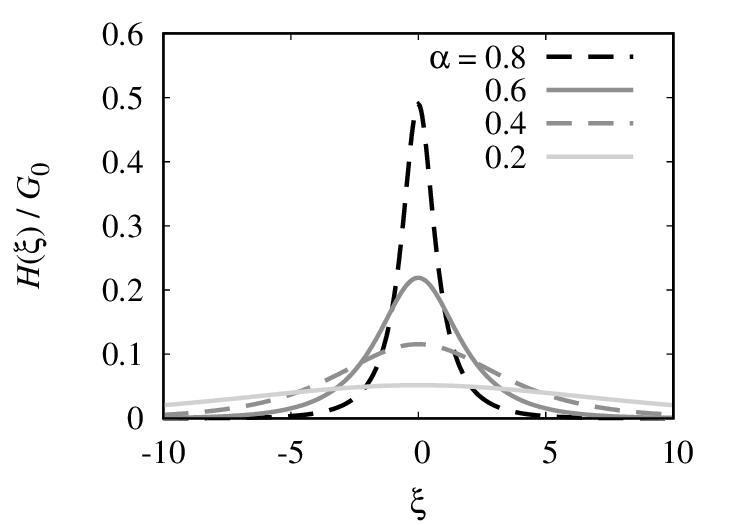}%
\end{center}
\caption{Relaxation spectra of the fractional Maxwell model with $\alpha = 0.2, 0.4, 0.6, 0.8$.
 (For $\alpha = 1$, the relaxation spectra becomes the Dirac delta function and thus not shown here.)
 \label{relaxaton_spectrum_fractional_maxwell}}
\end{figure}

Although eq~\eqref{relaxation_modulus_fractional_maxwell} is not simple,
the corresponding relaxation spectrum can be expressed in terms
of the elementary functions\cite{Mainardi-book,Carpinteri-Mainardi-book}:
\begin{equation}
 \label{relaxation_spectrum_xi_fractional_maxwell}
  H(\xi) = \frac{G_{0}}{2 \pi} \frac{\sin (\alpha \pi)}{\cosh(\alpha \xi) + \cos(\alpha \pi)}.
\end{equation}
Fig.~\ref{relaxaton_spectrum_fractional_maxwell} shows the relaxation
spectrum by eq~\eqref{relaxation_spectrum_xi_fractional_maxwell}.
We observe that $H(\xi)$ of the fractional Maxwell model has a single peak at $\xi = 0$ and is symmetric: $H(\xi) = H(-\xi)$.
(This symmetry gives $G''(\omega) = G''(\tau_{0}^{-2} \omega^{-1})$.)

For relatively small $\alpha$, the peak shape of $H(\xi)$ in Fig.~\ref{relaxaton_spectrum_fractional_maxwell}
seems to be similar to that of the Gaussian. Thus we expect that
$H(\xi)$ of the fractional Maxwell model can be approximated by that of the log-normal relaxation spectrum model.
We expand eq~\eqref{relaxation_spectrum_xi_fractional_maxwell} around $\xi = 0$ and approximate $H(\xi)$ as
\begin{equation}
 \label{relaxation_spectrum_xi_fractional_maxwell_expansion}
 \ln H(\xi) \approx \ln \left[\frac{G_{0}}{2 \pi} \frac{\sin (\alpha \pi)}{1 + \cos(\alpha \pi)}\right]
  - \frac{\alpha^{2}}{2 [1 + \cos (\alpha \pi)]} \xi^{2}.
\end{equation}
By comparing eq~\eqref{relaxation_spectrum_xi_fractional_maxwell_expansion}
with eq~\eqref{logarithmic_relaxation_spectrum_most_probable},
we find that the fractional Maxwell model can be approximated as the log-normal relaxation spectrum model:
\begin{equation}
 \label{relaxation_spectrum_xi_fractional_maxwell_approx}
  H(\xi) \approx \frac{G_{0,\text{eff}} }{\sqrt{2 \pi s_{\text{eff}}^{2}}}
  \exp\left( - \frac{\xi^{2}}{2 s_{\text{eff}}^{2}}\right),
\end{equation}
where $G_{0,\text{eff}}$ and $s_{\text{eff}}$ are given as
\begin{align}
 \label{effective_relaxation_intensity_fractional_maxwell}
 G_{0,\text{eff}} & 
 = \frac{G_{0}}{\sqrt{2 \pi}} \frac{\sin (\alpha \pi)}{\alpha \sqrt{1 + \cos(\alpha \pi)}}  , \\
 \label{effective_standard_deviation_fractional_maxwell}
 s_{\text{eff}} & = \frac{\sqrt{1 + \cos(\alpha \pi)}}{\alpha}.
\end{align}
From eq~\eqref{effective_standard_deviation_fractional_maxwell},
$s_{\text{eff}} = 0$ for $\alpha = 1$. Thus the Maxwell model is correctly recovered,
as expected. Also, at the limit of $\alpha \to 0$, $s_{\text{eff}}$ diverges ($s_{\text{eff}} \to \infty$).

\begin{figure}[tb]
\begin{center}
 \includegraphics[width=1.\refwidth]{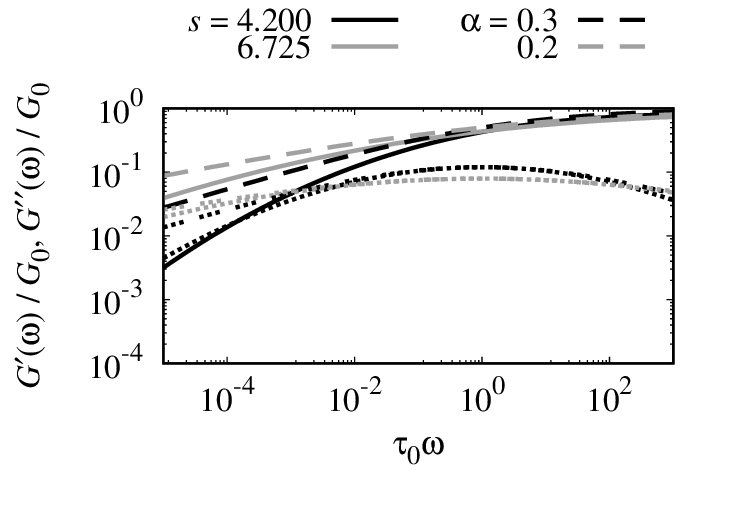}%
\end{center}
\caption{Comparison of the storage and loss moduli of the fractional Maxwell
 model (solid and dotted curves) and the log-normal relaxation spectrum model
 (dashed and dot-dashed curves).
 $\alpha = 0.2$ and $0.3$ for the fractional Maxwell model, and 
 $s$ and $G_{0}$ for the log-normal relaxation spectrum model
 are calculated by eqs~\eqref{effective_relaxation_intensity_fractional_maxwell}
and \eqref{effective_standard_deviation_fractional_maxwell}.
 \label{complex_modulus_comparison_log_normal_fractional_maxwell}}
\end{figure}

By using eq~\eqref{effective_standard_deviation_fractional_maxwell},
we can estimate the standard deviation of the log-normal relaxation model
from the exponent of the fractional Maxwell model.
Of course, the fractional Maxwell model does not coincide to the
log-normal relaxation model. The correspondence between two models
is justified only in the limited time or frequency range.
As an example, we show storage and loss moduli of the fractional Maxwell model and
those of the log-normal relaxation spectrum model with parameters $s_{\text{eff}}$ and $G_{0,\text{eff}}$
estimated by eqs~\eqref{effective_relaxation_intensity_fractional_maxwell}
and \eqref{effective_standard_deviation_fractional_maxwell}, in Fig.~\ref{complex_modulus_comparison_log_normal_fractional_maxwell}.
We observe that the loss moduli of two models agree very well around $\tau_{0} \omega = 1$.
The log-normal relaxation model may be used as an approximation for the fractional Maxwell model around the
peak of the loss modulus, and vice versa.

\subsection{Comparison with Experimental Data}
\label{comparison_with_experimental_data}

In this subsection, we attempt to utilize the log-normal relaxation
spectrum model to analyze some experimental rheology data.
Here we employ storage and loss moduli data of
high-density polyethylene (HDPE, Sigma-Aldrich, 547999) in melt and solid states.
The details of the experiments are summarized in Appendix~\ref{experimental}.

The molecular weight distribution of a commercial HDPE is typically broad.
As a result, often we cannot observe the clear terminal behavior at the low-frequency region.
Fulchiron and coworkers\cite{Fulchiron-Michel-Verney-Roustant-1995}
reported that the storage and loss moduli of some commercial polypropylenes
can be fitted to the log-normal relaxation spectrum model\cite{Fulchiron-Michel-Verney-Roustant-1995}.
We expect that the storage and loss moduli of an HDPE melt can be also expressed well
as the log-normal relaxation spectrum model.

\begin{figure}[tb]
\begin{center}
 \includegraphics[width=1.\refwidth]{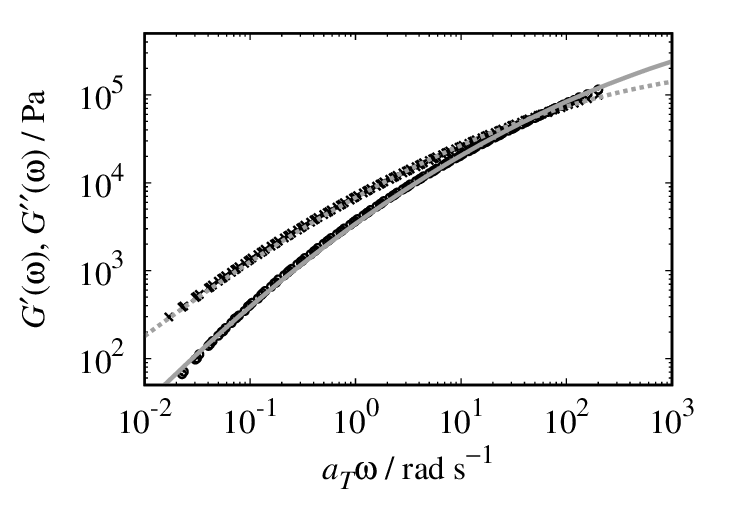}%
\end{center}
\caption{Storage and loss moduli of an HDPE melt at the reference temperature $T_{\text{ref}} = 140^{\circ}\text{C}$ (circles and crosses) and the log-normal
 relaxation spectrum model (solid and dotted curves).
 The parameters for the log-normal relaxation spectrum model are
 $G_{0} = 1.1 \times 10^{6}\,\text{Pa}$, $\tau_{0} = 6.5 \times 10^{-5}\,\text{s}$,
 and $s = 3.4$.
 \label{complex_modulus_comparison_log_normal_hdpe_melt}}
\end{figure}

Fig.~\ref{complex_modulus_comparison_log_normal_hdpe_melt} shows the
storage and loss moduli of an HDPE melt, and those by the log-normal relaxation spectrum model.
The experimental data are master curves constructed with the data at several different temperatures
by the time-temperature superposition.
The reference temperature for the master curves is $T_{\text{ref}} = 140^{\circ}\text{C}$.
The parameters in the log-normal relaxation spectrum model
are tuned manually. The results are $G_{0} = 1.1 \times 10^{6}\,\text{Pa}$, $\tau_{0} = 6.5 \times 10^{-5}\,\text{s}$,
and $s = 3.4$.
Although the agreement is not perfect, we observe that the broad relaxation
mode distribution is well captured by the log-normal relaxation model.
By using eq~\eqref{second_moment_average_relaxation_time}, the second-moment
average relaxation time is estimated as $\langle \tau \rangle_{w} = 2.2 \times 10^{3}\,\text{s}$.
Considering the fact that the terminal behavior is not observed in Fig.~6,
this estimate seems to be reasonable.

\begin{figure}[tb]
\begin{center}
 \includegraphics[width=1.\refwidth]{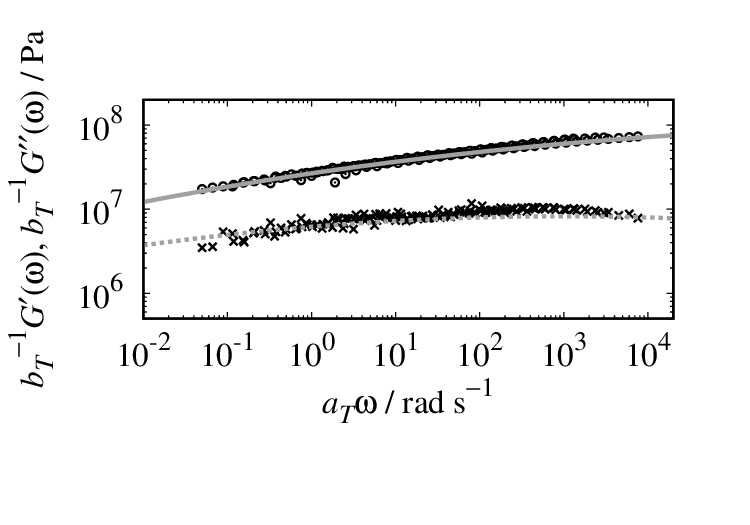}%
\end{center}
\caption{Storage and loss moduli of an HDPE solid at the reference temperature $T_{\text{ref}} = 100^{\circ}\text{C}$ (circles and crosses) and the log-normal
 relaxation spectrum model (solid and dotted curves).
 The parameters for the log-normal relaxation spectrum model are
 $G_{0} = 1.2 \times 10^{8}\,\text{Pa}$, $\tau_{0} = 1.0 \times 10^{-3}\,\text{s}$,
 and $s = 9.0$.
 \label{complex_modulus_comparison_log_normal_hdpe_solid}}
\end{figure}

HDPE is a typical crystalline polymer and form crystalline structures when
cooled and solidified from a melt.
Crystalline polymer solids exhibit very broad relaxation time distributions\cite{Li-Petzold-Ranga-Yu-vanNiekerk-ThurnAlbrecht-Men-2025},
which reflect motions of polymer chains in crystalline lamellar superstructures.
It is not easy to characterize their rheological behaviors
from the frequency-dependence of the storage and loss moduli.

Fig.~\ref{complex_modulus_comparison_log_normal_hdpe_solid}
shows the storage and loss moduli of an HDPE solid, and the 
those by the log-normal relaxation spectrum model.
The experimental data are master curves constructed by the time-temperature superposition,
and the reference temperature is $T_{\text{ref}} = 100^{\circ}\text{C}$.
Both the storage and loss moduli weakly depend on the frequency,
and we cannot observe the terminal behavior in the examined angular
frequency range.
As before, the parameters in the log-normal relaxation spectrum model
are tuned manually, and we have $G_{0} = 1.2\times10^{8}$, $\tau_{0} = 1.0\times 10^{-3}$, and $s = 9.0$.
The value of $s$ is much larger than that for a melt.
(Because $s$ is large, the fractional Maxwell model may be utilized as well.
From eq~\eqref{effective_standard_deviation_fractional_maxwell},
the exponent of the corresponding fractional Maxwell model will be $\alpha = 0.15$.)
Although the log-normal relaxation spectrum model does not perfectly reproduce
the experimental data, it reasonably mimics a very broad relaxation time
distribution. 
The second-moment average relaxation time can be estimated
as $\langle \tau \rangle_{w} = 5.8 \times 10^{49}\text{s}$.
The relaxation
time is extremely long and practically it is impossible to observe
by experiments.
This result is consistent with our naive expectation.

\section{CONCLUSIONS}
\label{conclusions}

We analyzed the relaxation modulus by using information geometry
and information theory. We showed that the distance between
probability densities with different life times becomes
the Euclidean distance, if we employ
the logarithmic relaxation time $\xi = \ln (\tau / \tau_{0})$.
This result theoretically supports the use of $H(\xi)$, instead of $h(\tau)$,
as the most natural and suitable expression.
We also showed that the most probable distribution for $\xi$ is
the normal distribution, and thus the log-normal relaxation spectrum
model is the information-theoretically most probable relaxation model.
We analyzed the properties of the log-normal relaxation spectrum model,
and compared it with the fractional Maxwell model.
We applied the log-normal relaxation spectrum model
to analyze the experimental data of an HDPE both at melt and solid states,
and showed that the log-normal relaxation spectrum model can reasonably
explain the linear viscoelasticity data.
Information geometry and information theory are currently widely used in
the field of data science, but not in the field of rheology.
We expect that information-theoretical analyses will provide some new
aspects of rheology.

\section*{ACKNOWLEDGMENT}

The author thanks Prof.~Ryuichi Tarumi (Osaka University) and
Dr.~Frank Nielsen (Sony Computer Science Laboratories) for giving an opportunity to study information geometry.
This work was supported by Grant-in-Aid (KAKENHI) for Scientific Research
Grant B No.~JP23H01142.

\appendix

\section*{APPENDIX}

\section{Experimental}
\label{experimental}

\subsection{Materials and Methods}
\label{materials_and_methods}

A high-density polyethylene (HDPE, Sigma-Aldrich, 547999) is used as received.
A controlled-stress rheometer (discovery hybrid rheometer, DHR-2, TA Instruments) is
used to measure the storage and loss moduli of HDPE at various temperatures,
both at melt and solid states.

For high temperature measurements above the melting point, 
a parallel-plate fixture with a diameter $25\,\text{mm}$ is used and the gap size
is set to $1\,\text{mm}$. Oscillatory strains with a strain amplitude $1\%$ is applied
to the sample and the storage and loss moduli, $G'$ and $G''$, are measured.
The angular frequency range is $0.05\,\text{rad}/\text{s} \le \omega \le 200\,\text{rad}/\text{s}$,
and the measurement temperatures are $T = 140, 160, 180,$ and $200^{\circ}\text{C}$.

For low temperature measurements below the melting point,
a parallel-plate fixture with a diameter $8\,\text{mm}$ is used.
The molten sample is loaded between the plates and the gap size is set to $1\,\text{mm}$. 
The sample was equilibrated at $140^{\circ}\text{C}$ (above the melting temperature) and then cooled to $100^{\circ}\text{C}$ 
(below the melting temperature) by $0.5\text{K}/\text{min}$.
During the solidification,
the density of the sample changes largely and it is impossible to keep the initial gap size.
Therefore, the axial force for the plate is controlled to be $0\,\text{N}$ while the temperature change,
and the gap is allowed to change freely.
When the temperature of the sample becomes the measurement temperature
and the sample is sufficiently relaxed, the applied axial force is changed to $10\text{N}$ 
(which corresponds to the normal stress $4.0 \times 10^{5}\text{Pa}$) in order to prevent delamination and slippage,
and the storage and loss moduli, $G'$ and $G''$ are measured.
To improve the accuracy, oscillatory stresses are applied to the sample,
instead of the oscillatory strains. The creep compliance $J'$ and
$J''$ are measured and then converted to $G'$ and $G''$.
We set the stress amplitude as $9.9 \times 10^{3}\,\text{Pa}$.
The angular frequency range is $0.05\,\text{rad}/\text{s} \le \omega \le 200\,\text{rad}/\text{s}$,
and the measurement temperatures are $T = 100$, $90$, $80$, $70$, and $60^{\circ}\text{C}$.
(The axial force is controlled to be $0\,\text{N}$ when the temperature is changed.)

\begin{figure}[tb]
\begin{center}
 \includegraphics[width=1.\refwidth]{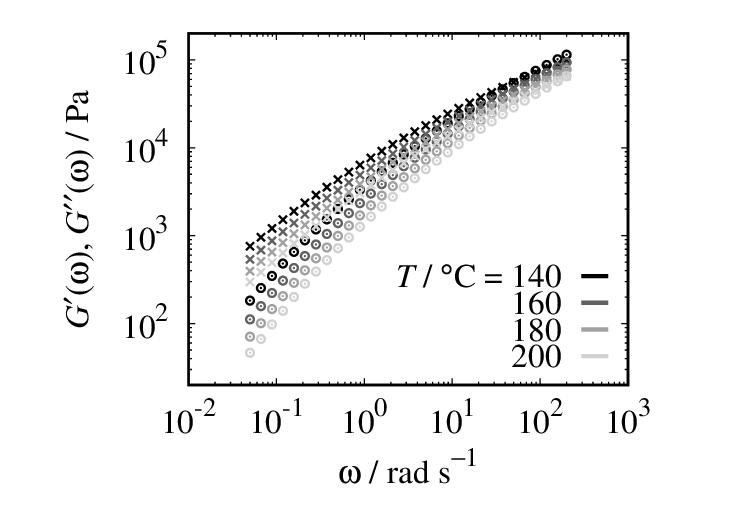}%
\end{center}
\caption{Storage and loss moduli of an HDPE melt at 
 $T = 140, 160, 180,$ and $200^{\circ}\text{C}$.
 Circles and crosses represent storage and loss moduli data, respectively.
 \label{complex_modulus_comparison_hdpe_melt}}
\end{figure}

\begin{figure}[tb]
\begin{center}
 \includegraphics[width=1.\refwidth]{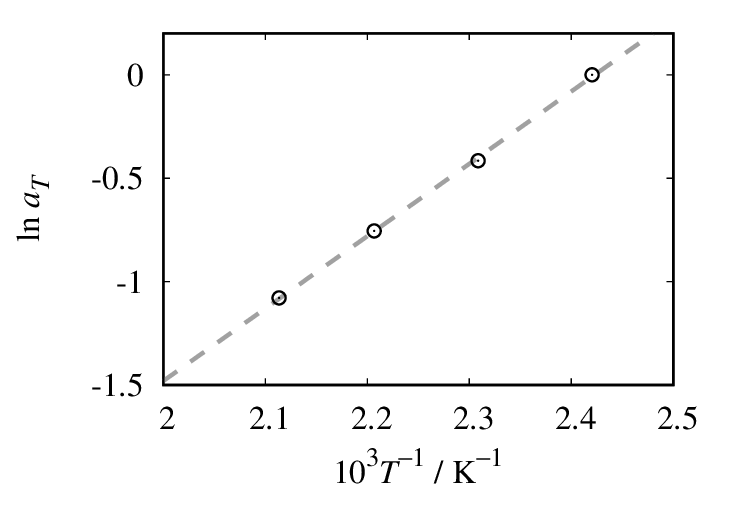}%
\end{center}
\caption{The Arrhenius plot for the horizontal shift factor $a_{T}$ of an HDPE melt.
 The reference temperature is $T_{\text{ref}} = 140^{\circ}\text{C}$.
 Symbols show the experimental data and the dashed line shows the fitting
 result to the Arrhenius form.
 \label{shift_factor_hdpe_melt}}
\end{figure}

\begin{figure}[tb]
\begin{center}
 \includegraphics[width=1.\refwidth]{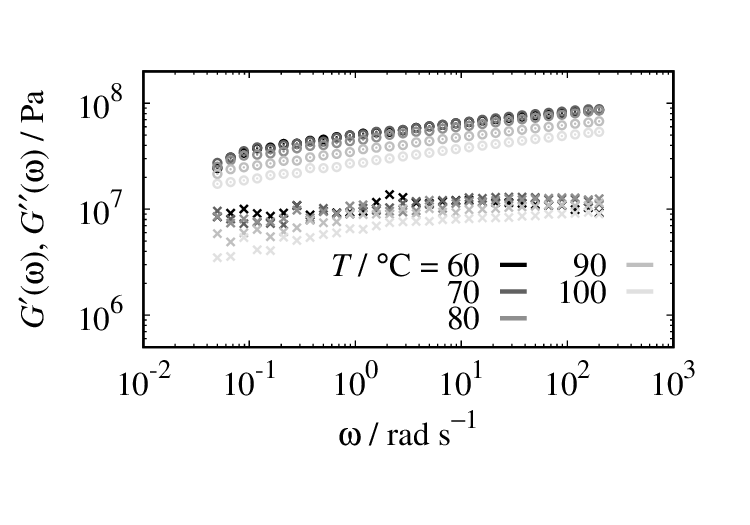}%
\end{center}
\caption{Storage and loss moduli of an HDPE melt at 
 $T = 60, 70, 80, 90,$ and $100^{\circ}\text{C}$.
 Circles and crosses represent storage and loss moduli data, respectively.
 \label{complex_modulus_comparison_hdpe_solid}}
\end{figure}

\begin{figure}[tb]
\begin{center}
 \includegraphics[width=1.\refwidth]{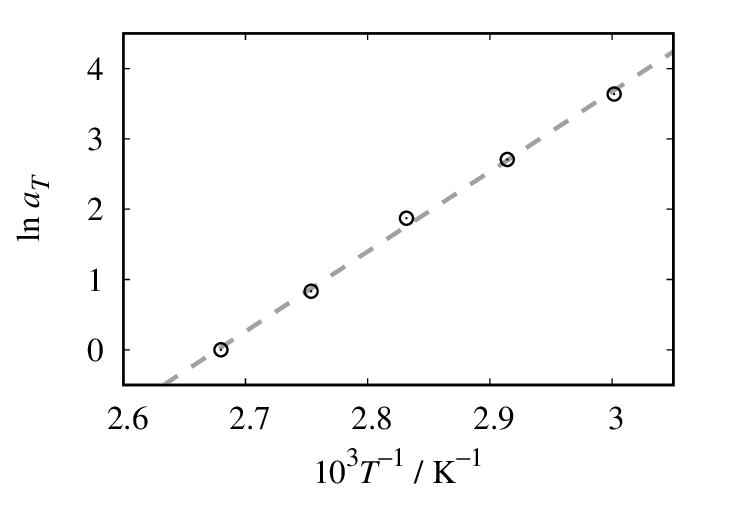}%
\end{center}
\caption{The Arrhenius plot for the horizontal shift factor $a_{T}$ of an HDPE solid.
 The reference temperature is $T_{\text{ref}} = 100^{\circ}\text{C}$.
 Symbols show the experimental data and the dashed line shows the fitting
 result to the Arrhenius form.
 \label{shift_factor_hdpe_solid}}
\end{figure}

\subsection{Results}
\label{results}

Fig.~\ref{complex_modulus_comparison_hdpe_melt} shows the
storage and loss moduli of an HDPE melt at several different temperatures.
Although the clear plateau is not observed, judging from the relaxation time,
HDPE is well-entangled. The relaxation time distribution seems to be
very broad.
By using the time-temperature superposition, we construct the master curves
for $G'(\omega)$ and $G''(\omega)$. We perform only the horizontal shift
by the horizontal shift factor $a_{T}$, and do not perform
the vertical shift. (The vertical shift factor is simply set as $b_{T} = 1$ for all the temperatures.)
We set the reference temperature as $T_{\text{ref}} = 140^{\circ}\text{C}$.
The storage and loss moduli data obtained by the time-temperature superposition
are shown in Fig.~\ref{complex_modulus_comparison_log_normal_hdpe_melt}.
The horizontal shift factor can be fitted to the Arrhenius form:
$\ln a_{T} = (E_{a} / R) (1 / T - 1 / T_{\text{ref}})$ where $E_{a}$ is 
the activation energy and $R$ is the gas constant.
Fig.~\ref{shift_factor_hdpe_melt} shows the Arrhenius plot for the
horizontal shift factor. 
The activation energy is estimated as $E_{a} = 2.9\times10^{1} \text{kJ}/\text{mol}$.
This value is comparable to typical activation energies of polyethylene melts\cite{Stadler-Kaschta-Munstedt-2008,Kida-Doi-Tanaka-Uneyama-Shiono-Masubuchi-2021}.

Fig.~\ref{complex_modulus_comparison_hdpe_melt} shows the
storage and loss moduli of an HDPE solid at several different temperatures.
We observe that both $G'(\omega)$
and $G''(\omega)$ depend on the angular frequency weakly. Such weak
angular frequency dependence can be attributed to a very broad relaxation
time distribution. The experimental data are scattered due to high
modulus of the sample and thus not that accurate.
We attempt to construct the master curves by using the time-temperature
superposition.
Unlike the case of the melt, we utilize both the horizontal and vertical
shift factors $a_{T}$ and $b_{T}$. (Note that the shift factors $a_{T}$ and $b_{T}$ are
different from those at the melt state, because the phase structures and
elementary relaxation processes of melt and solid samples are totally different.) By tuning $a_{T}$ and $b_{T}$,
we can construct the master curve shown in Fig.~\ref{complex_modulus_comparison_log_normal_hdpe_solid}.
Fig.~\ref{shift_factor_hdpe_solid} shows the Arrhenius plot for
the horizontal shift factor $a_{T}$.
The horizontal shift factor can be fitted to the Arrhenius form again,
and the activation energy is estimated as $E_{a} = 9.5\times 10^{1} \text{kJ}/\text{mol}$.
This value is comparable to typical activation energies of the crystalline relaxation
process (so-called the $\alpha$ relaxation) of polyethylene solids\cite{Strobl-book}.



\end{document}